\documentclass[]{article}
\usepackage{amsmath}
\usepackage{amssymb}
\usepackage{graphics}
\usepackage{graphicx}
\usepackage{epsfig}
\usepackage{dcolumn}
\usepackage{bm}
\usepackage{lineno}
\usepackage{ragged2e}

\title{Spark probability measurement of a single mask triple GEM detector}
\date{}

\begin{document}
\maketitle
\centering
{
\author {S.~Chatterjee$^a$,}
\author{U.~Frankenfeld$^b$,}
\author{C.~Garabatos$^b$,}
\author{J.~Hehner$^b$,}
\author{T.~Morhardt$^b$,}
\author{C.~J.~Schmidt$^b$,}
\author{H.~R.~Schmidt$^{b,c}$,}
\author{A.~Lymanets$^b$,}
\author{S.~Biswas$^{a,b,}$\footnote[1]{Corresponding Author:
saikat.ino@gmail.com, S.Biswas@gsi.de}}
}

\vspace*{0.5cm}
$^a${Department of Physics and Centre for Astroparticle Physics and Space Science
(CAPSS), Bose Institute, EN-80, Sector V, Kolkata-700091, India}
\vspace*{0.15cm}

$^b${GSI Helmholtzzentrum f{\"u}r Schwerionenforschung GmbH, Planckstrasse 1, D-64291 Darmstadt, Germany}
\vspace*{0.15cm}

$^c${Physikalisches Institut - Eberhard Karls Universit{\"a}t T{\"u}bingen, Auf der Morgenstelle 14, D-72076 T{\"u}bingen, Germany}

\vspace*{0.5cm}
\centering{\bf Abstract}
\justify

Triple Gas Electron Multiplier~(GEM) detectors will be used as a tracking device in the first two stations of CBM MUon CHamber~(MUCH), where the maximum particle rate is expected to reach $\sim$1~MHz/cm$^2$ for central Au-Au collisions at 8~AGeV. Therefore, the stable operation of the detector is very important. Discharge probability has been measured of a single mask triple GEM detector at the CERN SPS/H4 beam-line facility with a pion beam of $\sim$150~GeV/c and also in an environment of highly ionizing shower particles. The spark probability as a function of gain has been studied for different particle rates. The details of the experimental setup, method of spark identification and results are presented in this paper.    

\vspace*{0.25cm}

\section{Introduction}\label{intro}
The Compressed Baryonic Matter~(CBM)~\cite{CBM} experiment at the future Facility for Antiproton and Ion Research (FAIR)~\cite{FAIR} in Darmstadt, Germany, will explore the QCD phase diagram at low temperature and moderate to high baryonic density regime~\cite{CBMBOOK}. The decay of charmonium ($J/\psi$), low mass vector mesons $\rho^0$, $\omega^0$, $\phi^0$ in the muonic decay channel, i.e., $\mu^+\mu^-$ will be used as a probe to get an idea about the in-medium modifications of the particles~\cite{dimuon}. They will carry the information about the medium formation, the transition from the hadronic phase to the QGP phase and chiral symmetry restoration. 

The MUon CHamber~(MUCH) at CBM will be used dedicatedly for muon tracking~\cite{CBMdetector}. Since the product of multiplicity and branching ratio for the muonic decay mode is very small ($\sim$$10^{-3} - 10^{-8}$), therefore it is necessary to go up in interaction rate to get a signal, well separated from the background. This is only possible with the application of advanced instrumentation, including fast detectors with very high rate handling capability and good position resolution. MUCH will consist of five absorber layers of thickness 60, 20, 20, 30, 100~cm respectively. The first absorber will be made up of 60 cm carbon. The rest of the absorbers will be made up of iron. In between the absorbers (termed as stations), three active detector layers will be placed. To handle high rate, the triple GEM~\cite{sauli_GEM} detector technology has been chosen for the first two stations, and RPC or straw-tube will be used for the rest of the stations in the MUCH detector system \cite{s_biswas_gem, s_biswas_gem_2, s_biswas_gem_3, sayak_gem, RPA16, RPA17, SR19, SC19, SR19Straw}.

From simulation it has been found that the particle rate in the first four stations will be 0.8~MHz/cm$^2$, 0.1~MHz/cm$^2$, 15~kHz/cm$^2$, 5.6~kHz/cm$^2$ respectively for central Au-Au collisions at 8~AGeV~\cite{ekata}. To operate the detectors for a long period without any discharge is an essential criterion for MUCH.

GEM is made up of a thin kapton foil of thickness 50~$\mu m$ with 5~$\mu m$ copper cladding on both sides of the foil. A large number of holes are etched on the kapton using the photolithographic technique~\cite{photolithography}. Depending on the photolithographic technique used, the GEM foils can be divided into two types namely single mask and double mask GEM foils. In double mask technology for etching, the exposure of the metallized polymer foils, coated with a photosensitive resin, to ultra-violet light through masks from both sides of the sheet is required~\cite{CA02}. On the other hand, in single mask technology, following the masking, the metal and kapton are etched from one side~\cite{MV11, MA10}. The foil is first chemically etched to remove about half of the metal, opening the holes on the bottom side; a second kapton etching allows one to realise quasi-conical holes~\cite{FS16}.

The spark probability measurement of a double mask triple GEM detector has been done and reported earlier~\cite{sbiswas_spark}. Since the CBM GEM detectors will be of single mask type because of its large size, it is very important to measure the spark probability of a single mask GEM detector. 

The main goal of this study is to measure the spark probability of a single mask triple GEM detector with a high momentum pion beam and also for a heavy shower environment. A single mask triple GEM detector has been tested at the CERN SPS/H4 beamline facility with a pion beam of $\sim$$150$~GeV/c. In this test beam, the pulse height distribution from the detector, currents from the GEM foils and count rates from the detectors have been measured. The details of the spark identification, the value of spark probability, and its variation as a function of gain will be presented in this article.

\section{Description of the GEM module}\label{setup}

A single mask triple GEM detector having the dimension of (10~$\times$~10)~cm${^2}$ has been used during the beam test. The GEM foils with hole diameter of 70~$\mu$m, and pitch of 140~$\mu$m have been obtained from CERN. The drift gap, 2 transfer gaps, and induction gap have been kept at 3~mm, 2~mm, and 2~mm, respectively (3-2-2-2 configuration). A protection resistance of 10~M$\Omega$ has been employed to the top plane of each GEM foil and to the drift plane. Fig.~\ref{module} shows the schematic diagram of the GEM module, used in the test beam campaign. The read-out plane consist of 512 pads of (4~$\times$~4)~mm$^2$ size. All the readout pads have been routed to 4 connectors of 128 pins each. Even though the read-out plane was segmented for the module, in this study the signals obtained from all the 128 pads are summed by a sum-up board and a single output has been fed to a charge sensitive preamplifier. (The sum-up board is a specially designed board having connection from 128 pin to a single LEMO. Signals coming from any of the 128 pads will reach to a single preamplifier via the LEMO connector.)

\begin{figure}[htb!]
\begin{center}
\includegraphics[scale=0.32]{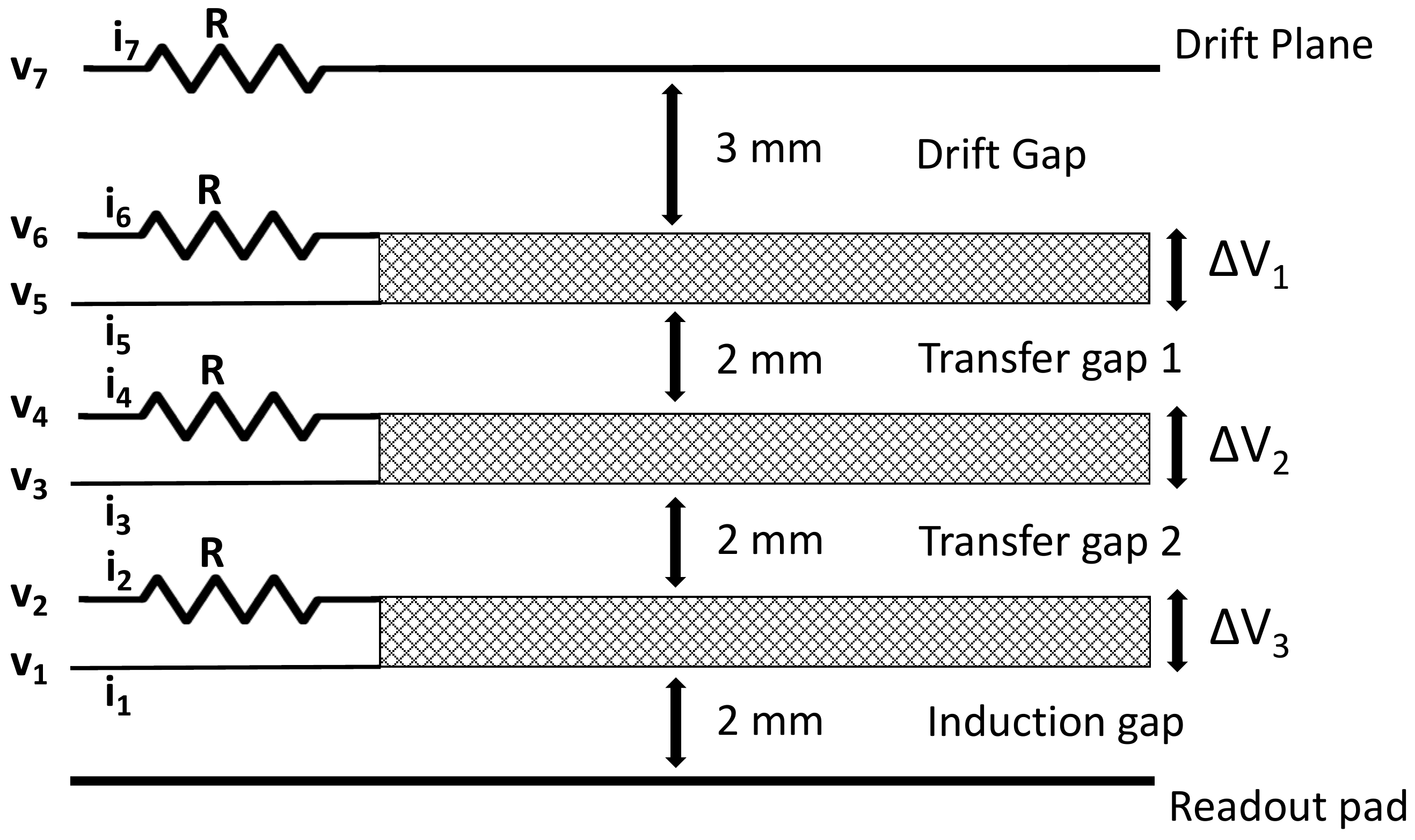}
\caption{\label{module} Arrangement of GEM foils, voltage and current distribution in different planes of the chamber.}\label{module}
\end{center}
\end{figure}

The analog signals from the preamplifiers have been put to the linear Fan-in-Fan-out (FIFO) module that gives four identical analog signal at the output exactly same as the input signal. For data acquisition, PXI LabView has been used~\cite{LABVIEW}. Signals from one output of the linear FIFO have been put to PXI LabView scope card for ADC spectra. Signals from another output have been fed to a NIM discriminator. The threshold to the signal has been set at 10~mV in the discriminator to eliminate noise. The discriminated signals have been counted using a PXI LabView scalar. The counts from the pad plane of the GEM detector are sampled for 100~ms binning. 

The GEM module has been operated throughout the experiment with a Ar/CO$_2$ gas mixture in the 70/30 volume ratio. The high voltages (HV) to the different GEM planes have been applied by a seven-channel HVG210 power supply made by LNF-INFN~\cite{HV}. This module allows for powering and controlling the applied voltages of a triple GEM detector. The module communicates with peripherals via CAN bus. The HVG210 power supply comprises seven almost identical channels, each of them being able to produce a specified voltage level with a current reading and current limiting option.
The currents of all channels were recorded and used to determine the occurrence of a spark. The applied voltages and measured currents on each channel from the lower plane of lowest GEM foil up to the drift plane are named as V$_1$ to V$_7$ and i$_1$ to i$_7$, respectively. The details of the electric field in the drift, transfer, and induction gap for a particular voltage configuration are summarised in Table~\ref{tab:table1} \cite{s_biswas_gem_3}.

\begin{center}
\begin{table}[htb!]
\begin{center}
\caption{Typical potential differences and fields on the various gaps of a triple GEM chamber, operated with Argon and CO$_{2}$ in a 70/30 mixing ratio.} \label{tab:table1}
\vspace*{0.4cm}
\begin{tabular}{|c|c|c|c|} \hline
Gap Name & Gap &  Potential & Field \\ 
 & width (mm) & Difference (V) & (kV/cm) \\ \hline
Drift & 3 &  400  & 1.33 \\ \hline
Transfer 1 & 2 &  395  & 1.98 \\ \hline
Transfer 2 & 2 &  395  & 1.98 \\ \hline
Induction & 2 &  390  & 1.95 \\ \hline
\end{tabular}\\
\end{center}
\end{table}
\end{center}

\section{Experimental setup}\label{set_up}

\begin{figure}[htb!]
\begin{center}
\includegraphics[scale=0.42]{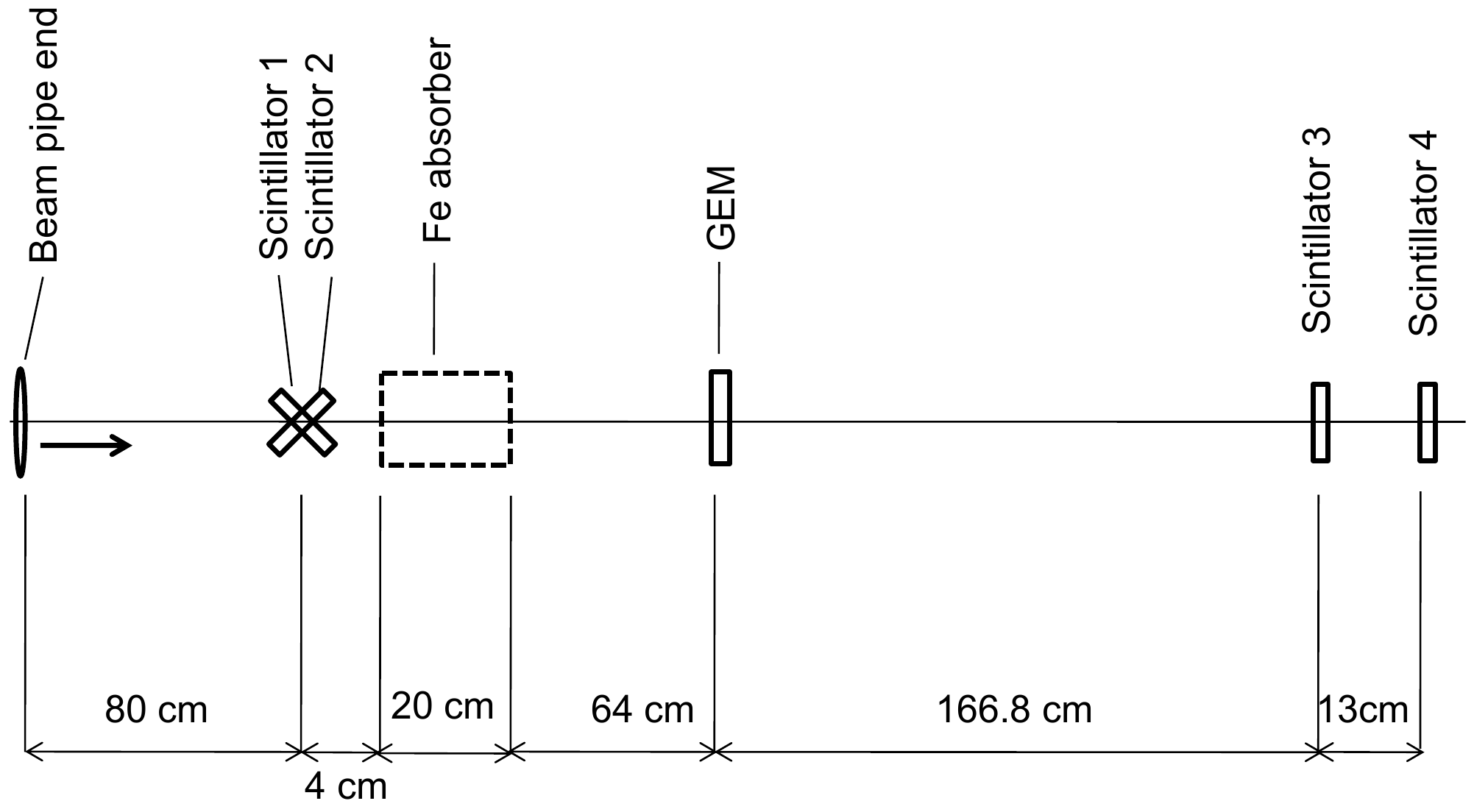}
\caption{\label{setup} A sketch of the experimental setup}\label{setup}
\end{center}
\end{figure}
 
The arrangement of the experimental setup used in the test beam campaign is shown in Fig.~\ref{setup}. Two crossed finger scintillators (Scintillator 1 and Scintillator 2, both having dimension (5~$\times$~5)~cm${^2}$) have been placed to monitor the incoming particle rate at a distance of 80~cm from the beam pipe end. The coincidence of those two scintillators (crossed area (5~$\times$~5)~cm${^2}$) have been used as a beam counter. The single mask triple GEM detector under test has been placed at a distance of 168~cm from the beam pipe end as shown in Fig.~\ref{setup}. An iron block of length 20~cm has been employed to generate a particle shower. The particle shower has been identified by the coincidence between the signals from the first two finger scintillators and no signal from the last two scintillators (Scintillator 3 and Scintillator 4, having dimension (10~$\times$~20)~cm${^2}$ and (20~$\times$~30)~cm${^2}$, respectively). The distance between the iron block and the GEM module was 64~cm as shown in Fig.~\ref{setup}. The centres of the 4 scintillators (Scintillator 1, 2, 3, 4), the iron block and the GEM module have been mechanically aligned with the centre of the beam pipe. FLUKA simulation package has been used to calculate the number of particles reaching on the detector surface after the shower production by the iron slab~\cite{fluka, fluka_ana, fluka_workshop}. From the FLUKA simulation the number of pion, neutron, muon, proton, kaon, and electron reaching on the GEM plane was found to be 2.4, 0.2, 0.009, 0.1, 0.3, 10, respectively, per primary pion of energy 150~GeV/c. In this study, to measure spark probability pion beam of rates 8, 27, 43, 48, 150 and 170~kHz have been used where as to produce shower, pion beam of rates 6, 50 and 120~kHz have been employed to hit a 20~cm thick iron block. The pion beam hit the GEM detector in an area of $\sim$~10~mm$^2$, where as in this set-up during shower, for each pion beam the number of secondary particles hitting the whole GEM plane of 100~cm$^2$ is 13.009 (sum of the numbers of secondary pion, neutron, muon, proton, kaon, and electron reaching GEM plane per primary pion).

The voltages and currents from all seven channels of the HVG210, counts from the scintillators, GEM detector and the pulse height of the GEM detector signals have been measured. In Section~\ref{res}, we shall discuss the relevance of those measurements for the investigation of the spark probability of the module.

\section{Results}\label{res}

In this test beam, the current in all the channels from the top and bottom plane of three GEM foils and the drift planes have been measured. The data for the ADC spectra have been stored for all voltage settings. The data for counts from the GEM detector and from the scintillators have been also stored. 

\subsection{ADC spectra}

\begin{figure}[htb!]
\begin{center}
\includegraphics[scale=0.47]{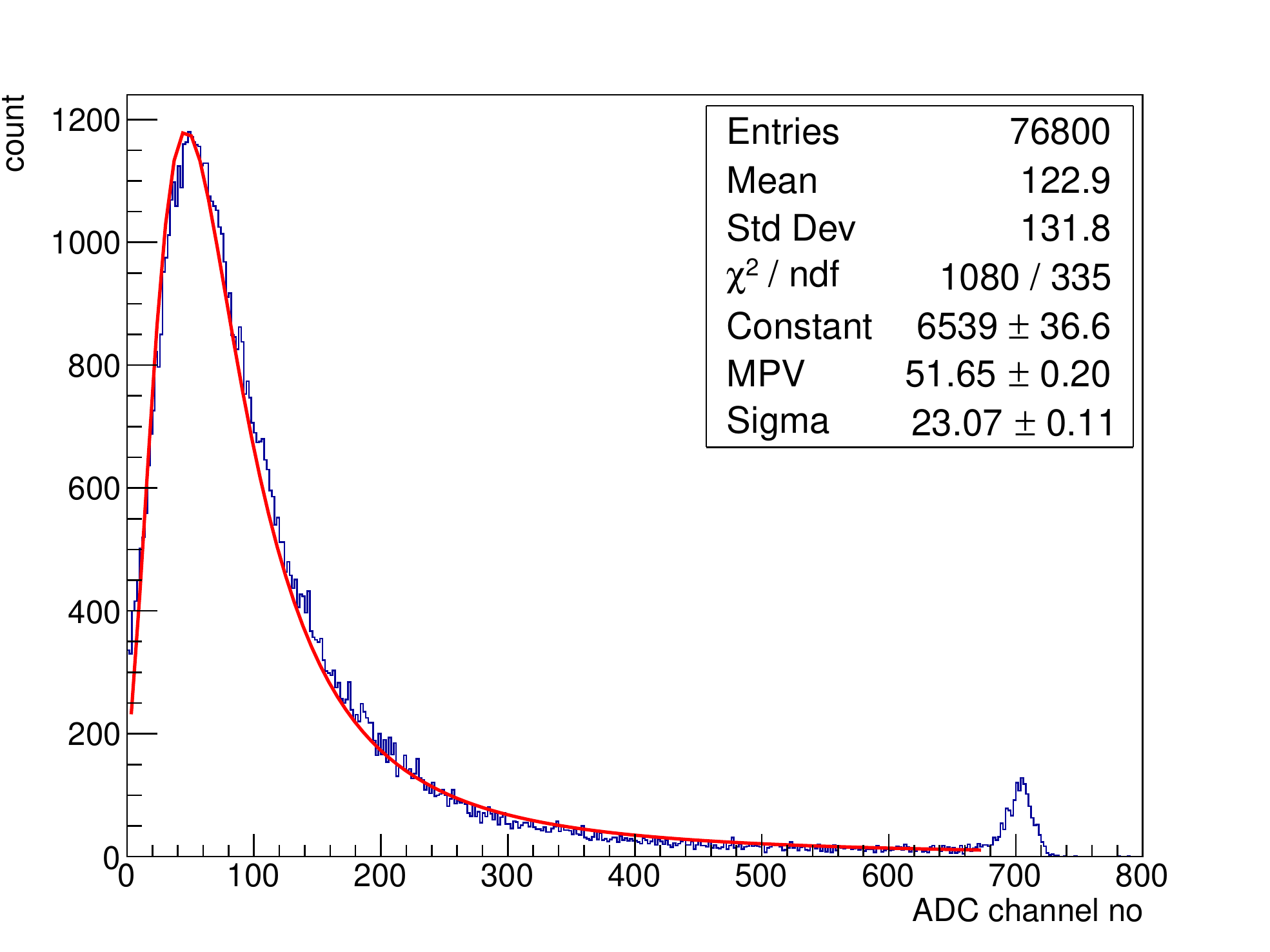}
\caption{(Colour online) ADC distribution for the pion of average rate 27~kHz with $\Delta V_1$=390~V, $\Delta V_2$=385~V and $\Delta V_3$=380~V and corresponding gain $\sim$~80000.}\label{pion}
\end{center}
\end{figure}

The ADC spectra of the detected particles have been studied. The ADC distribution for a pion beam of a typical average rate 27~kHz is shown in Fig.~\ref{pion}. The energy distribution of the minimum ionizing particle is expected to follow the Landau distribution~\cite{landau}, as observed from the ADC distribution for pion (Fig.~\ref{pion}) with a GEM voltage settings of $\Delta V_1$=390~V, $\Delta V_2$=385~V and $\Delta V_3$=380~V and corresponding gain $\sim$~80000. The Most Probable Value (MPV) of the distribution has been found at $\sim$~51 ADC channel and a small saturation peak has been observed at 700 ADC channel.

\begin{figure}[htb!]
\begin{center}
\includegraphics[scale=0.47]{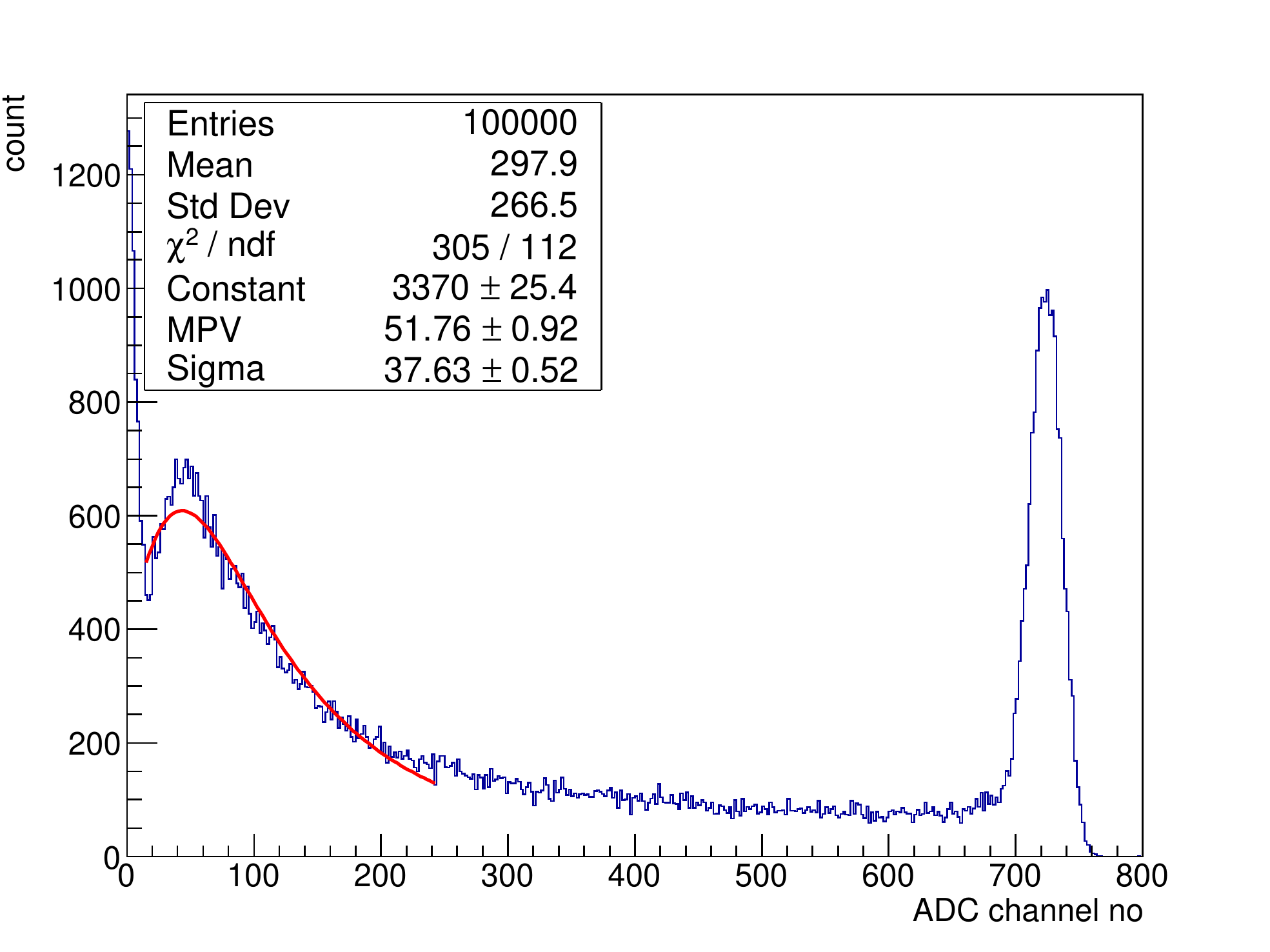}
\caption{(Colour online) ADC distribution for shower environment with $\Delta V_1$=390~V, $\Delta V_2$=385~V and $\Delta V_3$=380~V and corresponding gain $\sim$~80000.}\label{shower}
\end{center}
\end{figure}

The ADC distribution for the heavily ionizing particles produced after the shower is shown in Fig.~\ref{shower} with GEM voltage configuration of $\Delta V_1$=390~V, $\Delta V_2$=385~V and $\Delta V_3$=380~V and corresponding gain $\sim$~80000. The MPV of the distribution has also been found at $\sim$~51 ADC channel and a large saturation peak has been observed. Although it is somehow unexpected that the MPV is the same both for the heavily ionizing and minimum ionizing particles. The mean of the distribution for the pion beam and for shower, at the same voltage settings have been found to be at 122.9 and 297.9 ADC value, respectively. The average energy distribution by the particles produced in the shower is higher. For shower the mean value has increased only by a factor of $\sim$~2.5, which is also somehow surprising taking into account the steep increase of the Bethe-Bloch formula towards small velocities. The large saturation peak for the shower environment reflects the existence of heavily ionizing particles in the shower. In this work PXI LabView based scope card has been used to store the ADC spectra. It digitises the difference of the maximum and minimum edge of a signal and takes the number as the amplitude of the signal. For saturated signal also, although the maximum edge is more or less fixed but both the edges (maximum and minimum) can fluctuate a bit. Accordingly the digitised value also fluctuates. This is the reason for the broadening of the saturation peak in both Fig.~\ref{pion} and Fig.~\ref{shower}. From the scope data, it can be inferred that the detectors were in good condition during the beam time. 

\begin{figure}[htb!]
\begin{center}
\includegraphics[scale=0.44]{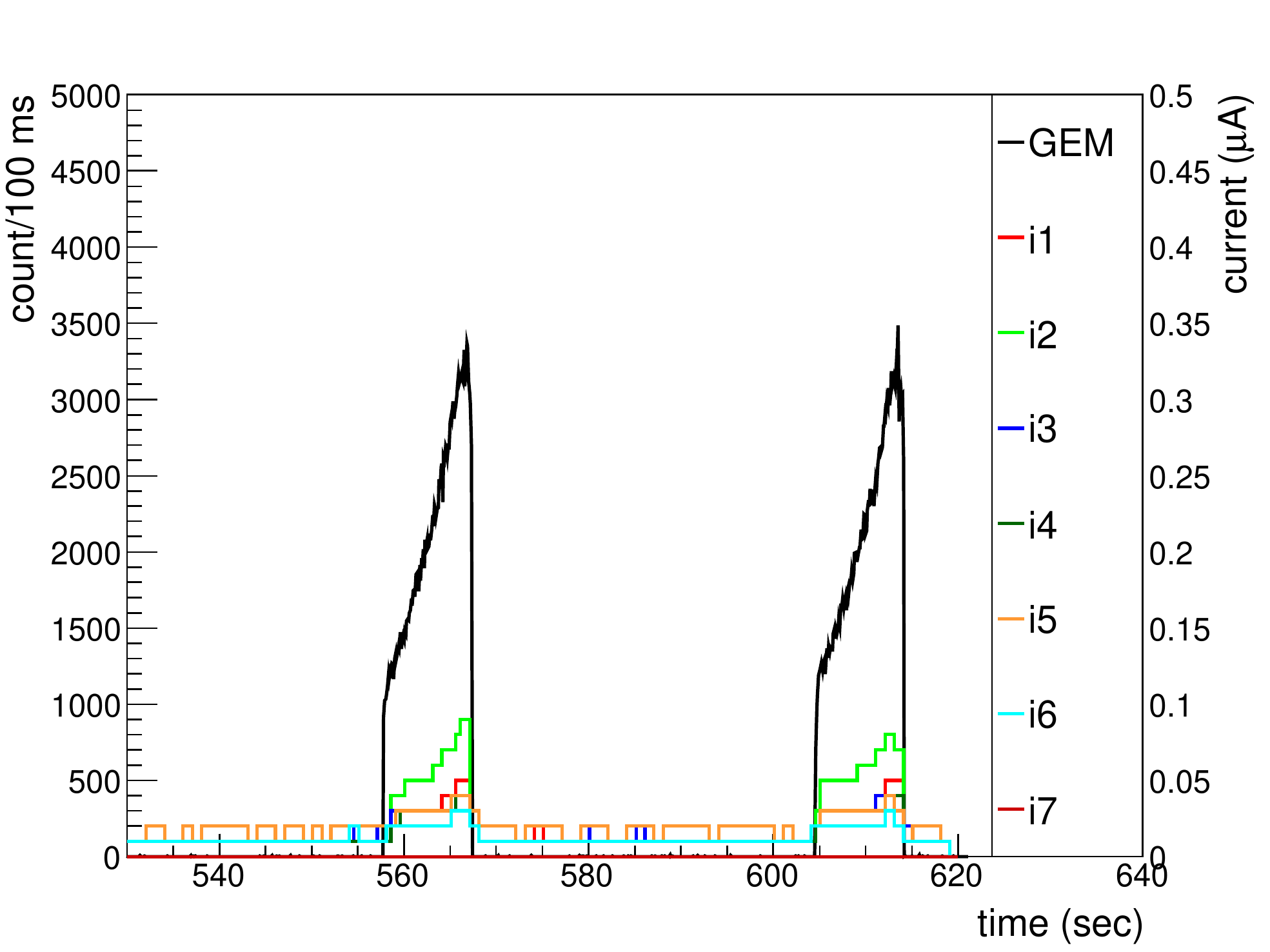}
\caption{(Colour online) Currents and the GEM counting rate: Pion beam 27 kHz. The GEM count rate is plotted in the units of counts/100~ms. The different currents i1 to i7 correspond to V1 to V7.}\label{pion_spill}
\end{center}
\end{figure}

\subsection{Measurement of current}

\begin{figure}[htb!]
\begin{center}
\includegraphics[scale=0.44]{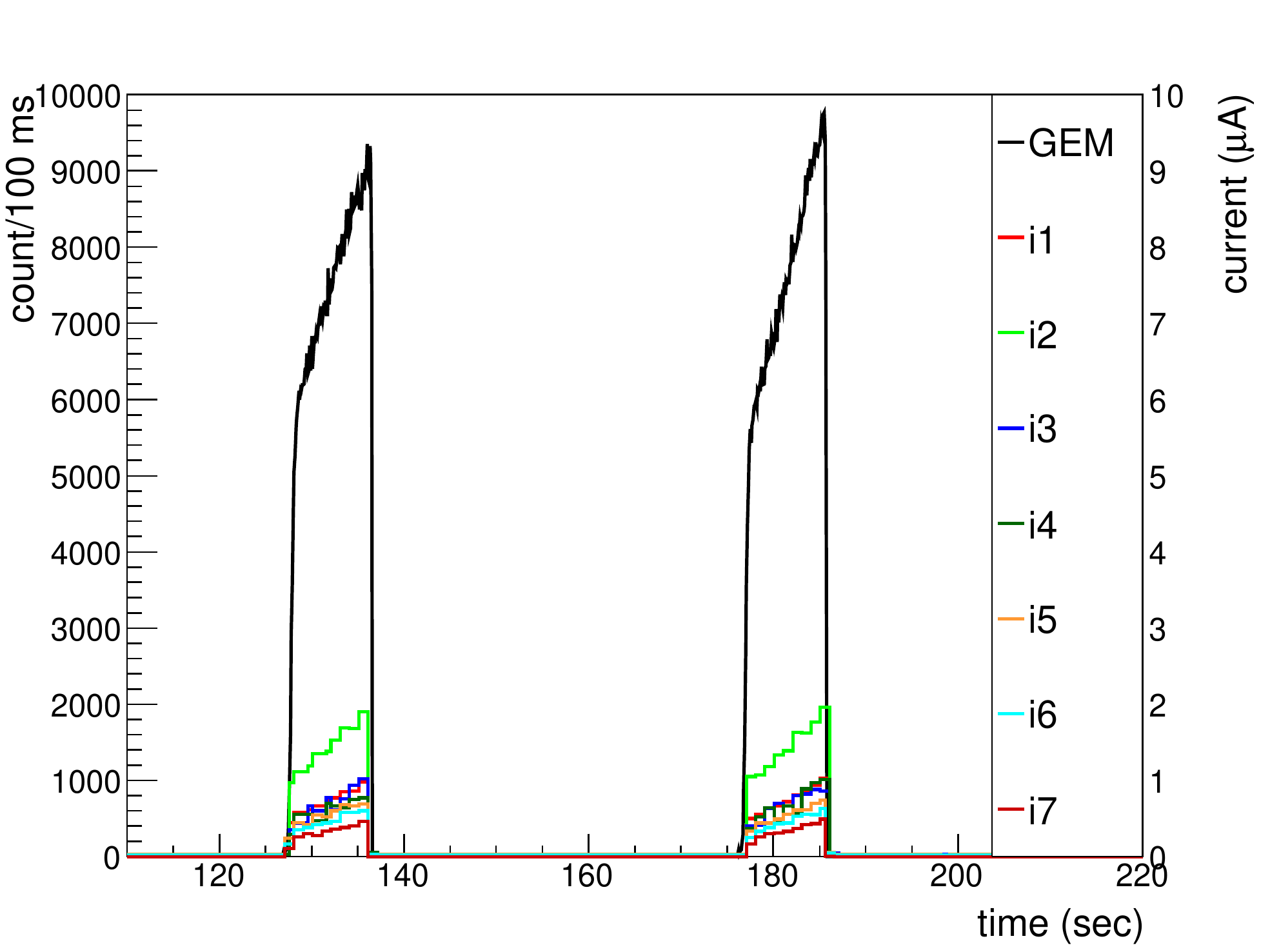}
\caption{(Colour online) Current and the GEM counting rate during Shower: Beam rate 120~kHz. The GEM count rate is plotted in the units of counts/100~ms. The different currents i1 to i7 correspond to V1 to V7.}\label{shower_spill}
\end{center}
\end{figure}

In this study, the currents from the drift plane, top and bottom plane of each of the GEM foils have been recorded using the HVG210~\cite{HV} high voltage power supply module. The counts from the GEM detector as well as from the scintillators are sampled for 100~ms binning. The variation in the current along with the count rate from the GEM module is shown as a function of time in Fig.~\ref{pion_spill} and Fig.~\ref{shower_spill}. 

Fig.~\ref{pion_spill} shows the variation of the currents and the GEM count rate during and in between the spills of the pion beam of average rate $\sim$27~kHz with GEM voltage settings of $\Delta V_1$=390~V, $\Delta V_2$=385~V and $\Delta V_3$=380~V and corresponding gain $\sim$~80000. In Fig.~\ref{shower_spill}, the variation in the currents and GEM count rate are shown for the shower produced by a pion beam of an average rate of 120~kHz hitting the 20~cm iron slab. The GEM voltage settings were $\Delta V_1$=385~V, $\Delta V_2$=380~V and $\Delta V_3$=375~V and corresponding gain was $\sim$~60000. The spill structure of the SPS beam increases with time reaches a maximum and then drops immediately to 0 as obtained from both the GEM detector and beam counter. The duration of the spill in the SPS beam is $\sim$10~s and the off spill time is $\sim$40~s. From Fig.~\ref{pion_spill} and Fig.~\ref{shower_spill}, the maximum absolute increase in current is observed in i2  i.e. on the top of the third GEM-foil, where the maximum number of ions reach.

\subsection{Measurement of spark probability}

The most important goal of this beam time was the measurement of spark probability. The spark probability is defined as the ratio of the number of sparks occurred in the detector and the total number of particles incident on it \cite{bachmann, Bencivenni, Alfonsi}. In this study, two different methods have been used to identify a spark in the GEM module as previously done for the double mask detector \cite{sbiswas_spark}. The first method identifies a spark if there is a sudden drop in the GEM counting rate. The second one determines a spark by the sudden jump in the current obtained from the top of each GEM foil.

\begin{figure}[htb!]
\begin{center}
\includegraphics[scale=0.44]{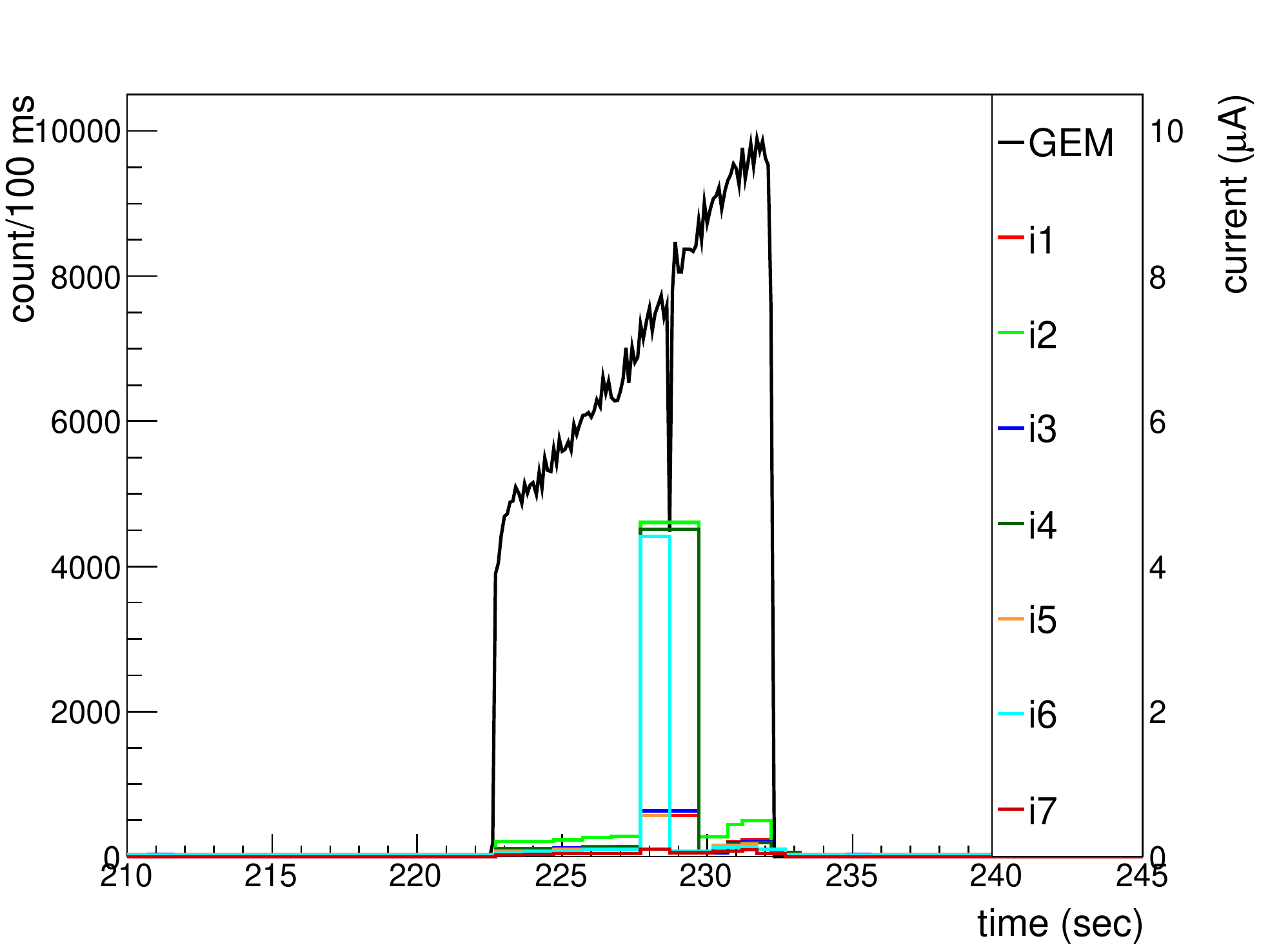}
\caption{(Colour online) Identification of spark from the drop in the GEM counting rate during a spill. In parallel, the currents on all GEM electrodes were registered and are displayed. The time axis is shown in the unit of second. i1 to i7 shown in different colours are the currents correspond to V1 to V7. The count rate, shown in black is in the unit of counts/100 ms.}\label{single_spark}
\end{center}
\end{figure}

\begin{figure}[htb!]
\begin{center}
\includegraphics[scale=0.44]{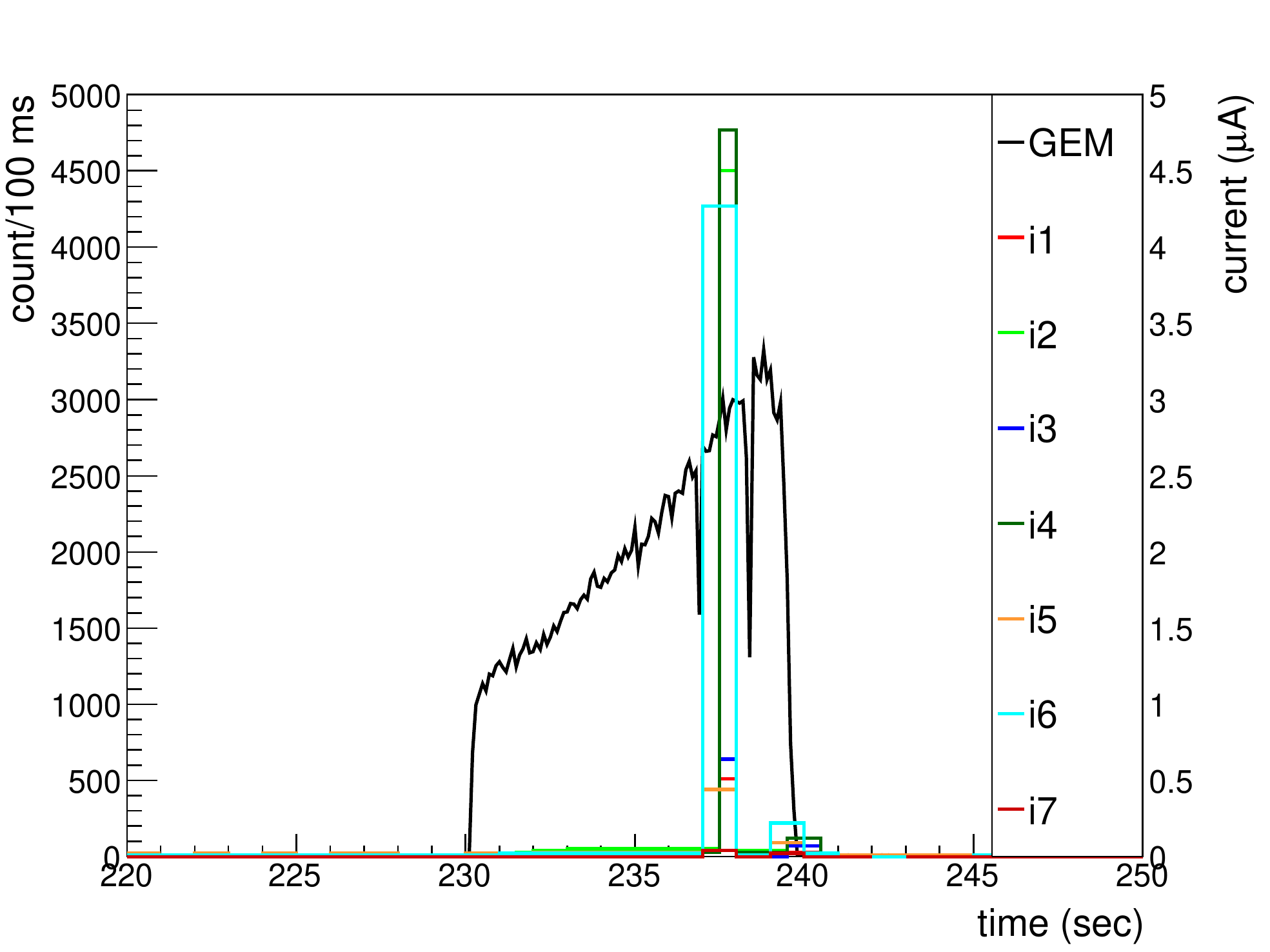}
\caption{(Colour online) Example of the spill where two sparks are observed. The time axis is shown in the unit of second. i1 to i7 shown in different colours are the currents correspond to V1 to V7. The count rate, shown in black is in the unit of counts/100 ms.}\label{double_spark}
\end{center}
\end{figure}

During a spark, the sudden drop of the electric field in the GEM hole reduces the gain of the detector, and as a result, the count rate of the chamber decreases. That is why it is a beneficial method to calculate the number of sparks that occurred in the GEM module during the spill. To identify a spark, the ratio of the counts from the GEM module and beam counter has been used. If the ratio drops below 65\% of its average value, then it is considered as a spark. Different threshold values between 50\% to 70\% have been tested, but no significant change in the result is observed. Above 70\%, the spark counts increase drastically because then all the small fluctuations in the count rate are considered as a spark, and below 50\% the spark count is coming to be zero. The above-mentioned definition has been used for the identification of spark in the analysis. In Fig.~\ref{single_spark}, the black line shows the count registered on the GEM module during a spill and a sudden drop in the count rate indicates the occurrence of spark in the chamber. Fig.~\ref{double_spark} shows that sometimes more than one spark is observed in the module during a spill.  

Measuring the currents from each layer of the GEM foil is another method of determining spark in the module. The current will increase particularly at the top of each GEM foil whenever there is a spark. In Fig.~\ref{single_spark} and Fig.~\ref{double_spark} the sudden jump in the current in the top layers of the GEM foil is observed when there is a drop in the GEM counting rate. The threshold for the current is set to 2 $\mu$A to define a spark, but the identification of the spark is more accurate if we use the first method i.e. from the drop in the GEM counting rate because of the sampling rate for the current monitoring is less than the sampling rate for the count data monitoring. If we consider Fig.~\ref{double_spark}, then the number of sparks is two if we count from the drop in the GEM counting rate, but it is coming to be one if we count the spark from the jump in the current. That is why, for our analysis, the spark probability is calculated from the drop in the counting rate of the GEM module during a spill.

\begin{figure}[htb!]
\begin{center}
\includegraphics[scale=0.47]{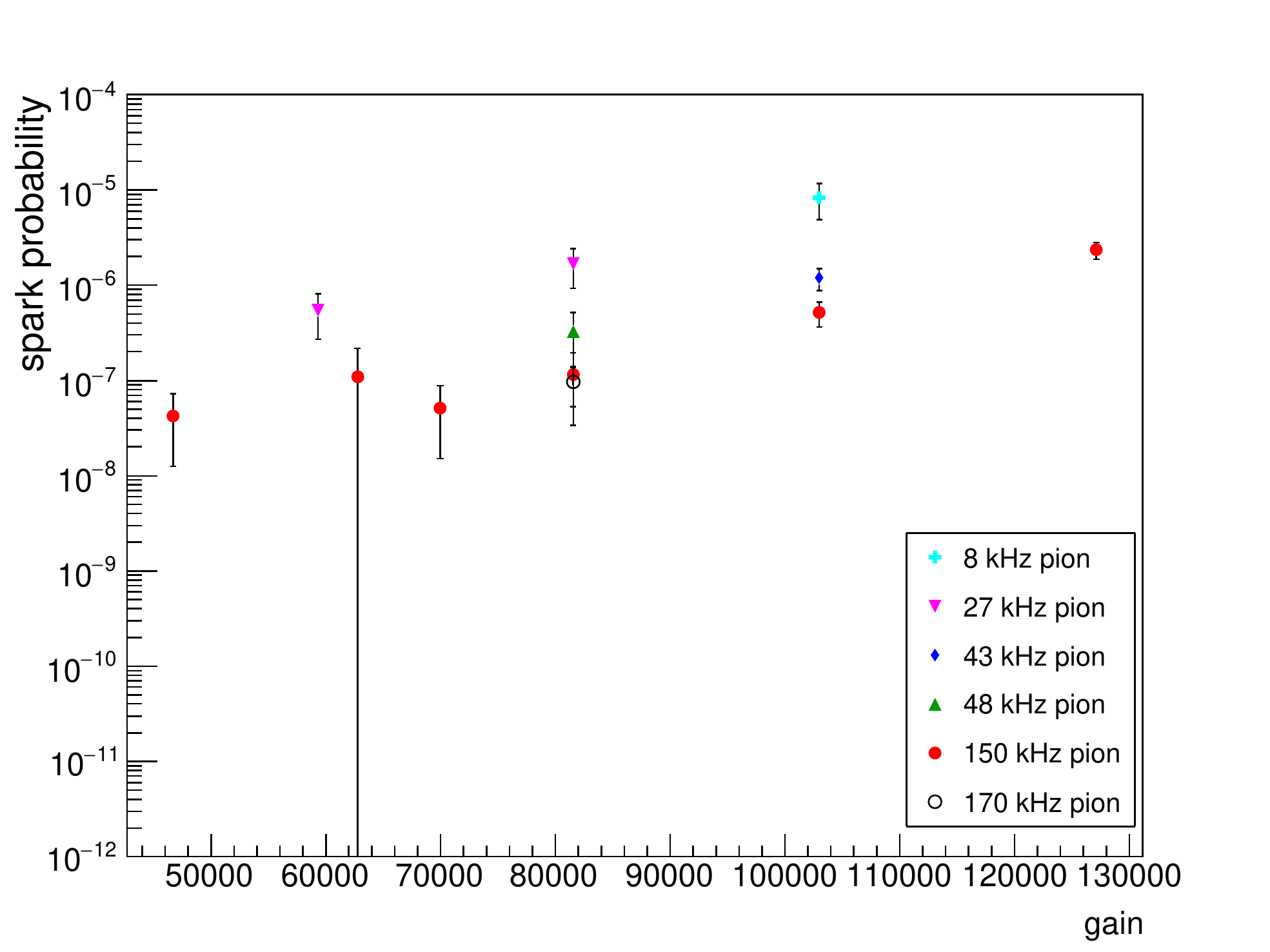}
\caption{(Colour online) Spark probability of the detector as a function of the gain.}\label{spark_prob}
\end{center}
\end{figure}

The calculated spark probability as a function of the gain of the module is shown in Fig.~\ref{spark_prob}. The gain of the module has been measured using a 5.9~keV Fe$^{55}$ X-ray source. During the beam time, the gain of the detector is found to be within the range of $\sim$~40000 to 130000 for the operational global GEM voltage ($\Delta V_1 + \Delta V_2 + \Delta V_3$) settings of 1120~V to 1185~V. In this operational global voltage range of 1120~V to 1185~V and corresponding gain between 40000 to 130000, taking 30 primary electrons per incident pion (minimum ionising particle) in the 3~mm drift gap the total number of electrons reaching readout will be 1.2~$\times$~10$^6$ to 3.9~$\times$~10$^6$, respectively. This corresponds to a total charge between 192~fC to 624~fC, respectively. The spark probability of the single mask triple GEM detector in "3-2-2-2 configuration" has been found to be $\sim$10$^{-7}$ for a 150 GeV/c pion beam of rate 150~kHz with a gas gain between 40000 and 80000. 

To calculate the spark probability of the GEM module in the shower environment, pion beam of rates 120, 50 and 6~kHz have been employed to a 20~cm thick iron slab. The  voltage settings of the detector were $\Delta V_1$=385~V, $\Delta V_2$=380~V and $\Delta V_3$=375~V respectively and the gain was $\sim$~60000. Total integrated number of pions incident on the iron slab for these three rates were 1.10~$\times$~10$^6$, 2.40~$\times$~10$^6$ and 3.32~$\times$~10$^5$, respectively. No spark has been detected in these settings using both the methods i.e. drop in GEM counting rate and jump in current. Though during the shower, the number of secondary particles on the detector surface (13.009 secondary particles reached the detector plane per pion) increases as we have seen from the FLUKA simulation, but still, no spark has been identified. 

\section{Summary and Outlook}

The spark probability of a single mask triple GEM detector in "3-2-2-2 configuration" has been measured with mostly pure pion beam and also for a shower produced by pion beam with a 20~cm thick iron block. Two different methods have been used to identify sparks in the chamber. In the first method the spark is identified by sudden drop in the GEM counting rate whereas in the second method it is done by the sudden jump in the current obtained from the top of each GEM foil. The variation of the spark probability as a function of the gain has been presented for the pion beam. For the pion beam the spark probability increases exponentially with the gain. The spark probability of the detector has been found to be $\sim10^{-7}$ for 150~GeV/c pion beam of rate 150~kHz with a gain between 40000 and 80000. No spark has been observed for shower produced by pion beams of rates of 120, 50 and 6~kHz after striking an iron slab of thickness 20~cm. The pion beam hit the chamber in an area of $\sim$~10~mm$^2$, where as in this set-up during shower, for each pion beam the number of secondary particles hitting the whole GEM plane of 100~cm$^2$ is 13.009. Consequently the particle density per unit surface area of the GEM detector is much smaller for the secondary particles produced in shower than that for the pion beam. In these measurements the particles hitting per unit area of the GEM detector for the pion beams of rate 8, 27, 43, 48, 150 and 170~kHz are $\sim$~ 0.8, 2.7, 4.3, 4.8, 15 and 17~kHz/mm$^2$, respectively and for that for shower produced by pion beam of rate 6, 50 and 120~kHz are $\sim$~ 0.008, 0.065 and 0.16~kHz/mm$^2$, respectively. That is the probable reason of not getting any spark in the shower set-up.

The comparison between the spark probability measurement for the double mask triple GEM detector as reported earlier~\cite{sbiswas_spark} and the present measurement for the single mask triple GEM detector are the followings. The drift gap of the double mask GEM chamber was 2~mm. In case of double mask triple GEM detector the spark probability was measured mostly for shower induced by a pion beam with a 10~cm thick iron absorber and also for a pure pion beam. In this operational global voltage range the gain of the detector was measured to vary between 20,000 and 50,000. 11~M$\Omega$ protection resistors were employed in all the seven channels. In this study the spark probability was found to be $\sim10^{-7}$ for 150~GeV/c pion beam and also for shower. On the other hand, the drift gap of the single mask GEM detector was 3~mm and the spark probability was measured mainly for pure pion beam of different rates and for shower produced by pion beams of rates of 120, 50 and 6~kHz after hitting an iron slab of thickness 20~cm. The detector was operated at gain between 40000 and 130000. A protection resistance of 10~M$\Omega$ has been employed only to the top plane of each GEM foil and to the drift plane. In this case the spark probability has been found to be $\sim10^{-7}$ for 150~GeV/c pion beam of rate 150~kHz with a gain between 40000 and 80000. No spark has been observed during the shower.

In this test beam the single mask triple GEM detector was operated at very high gain. Actually in CBM-MUCH the GEM chambers will be operated at a gain $\sim$~5000-8000. Extrapolating the value of spark probability for 150~GeV/c pion beam of rate 150~kHz it is coming $\sim10^{-9}$ at gain $\sim$~5000-8000. The value of the spark probability obtained from this beam test is little bit high for the operation of the CBM muon chambers at gain $\sim$~5000-8000. As an outlook, the measurements will be repeated in future and also at operational gain, with different electric fields in the Drift, Transfer and Induction gaps and with different value of current limiting protection resistor and will be communicated at a later stage.   

\section{Acknowledgements}

We are grateful to Dr. Anna Senger of GSI for the FLUKA simulation results. We would like to thank Jorrit C. L. Widder for his effort during the SPS test beam. We are thankful to Prof. Dr. Peter Senger of GSI and Dr. Subhasis Chattopadhyay of VECC, Kolkata for their support in course of this work.  We are also grateful to Dr. Leszek Ropelewski and Dr. Serge Duarte Pinto of RD51 for their valuable suggestions. S. Biswas acknowledges Dr. Ingo Fr{\"o}hlich of University of Frankfurt, Prof. Dr. Peter Fischer of Institut f{\"u}r Technische Informatik der Universit{\"a}t Heidelberg and the support of DST-SERB Ramanujan Fellowship (D.O. No. SR/S2/RJN-02/2012).
S. Chatterjee acknowledges his Institutional Fellowship research grant of Bose Institute. S. Chatterjee and S. Biswas would like to thank Ms. Shreya Roy, Prof. Sanjay K. Ghosh, Prof. Sibaji Raha, Prof. Rajarshi Ray, Prof. Supriya Das and Dr. Sidharth K. Prasad of Bose Institute for valuable discussions during the data analysis.

\end{document}